\documentclass{ws-procs9x6}

\begin{document}
\title{Statistical hadronization of charm in heavy ion collisions}
\author{A.P. Kostyuk}

\address{Institut f\"ur Theoretische Physik, Universit\"at  Frankfurt,
Germany\\
and\\
Bogolyubov Institute for Theoretical Physics,
Kyiv, Ukraine}

\maketitle

\abstracts{
 Production of open and hidden charm hadrons in heavy ion collisions is
considered within the statistical coalescence model (SCM). Charmed
quark-antiquark pairs are assumed to be created at the initial stage
of the reaction in hard parton collisions.
The number of these pairs is conserved during the evolution of the system.
At hadronization, the charmed (anti)quarks are distributed among open and
hidden charm hadrons in accordance with laws of statistical mechanics.
Important special cases: a system with a small number of charmed 
quark-antiquark pairs and charm hadronization in a subsystem of the 
whole system are considered.
   The model calculations are compared with the preliminary PHENIX data for
J/psi production at RHIC. Possible influence of the in-nuclear modification
of the parton distribution functions (shadowing) on the SCM results is studied.
}

The thermal hadron gas (HG) model explains successfully the chemical 
composition of light hadrons produced by colliding nuclei in a
wide range center-of-mass energies\cite{HG}. A large body of
experimental data can be well fitted with only three\footnote{Sometimes 
one more parameter, the strangeness suppression factor $\gamma_s$ is
introduced.} free
parameters: the temperature $T$, baryonic chemical potential
$\mu_b$ and volume $V$ at the point of chemical freeze-out. This
suggested the idea\cite{GG} to see, whether this model can be
applied also to charm particles. A straightforward generalization,
when particles containing heavy (anti)quarks are treated in the
same way as light hadrons, could hardly be successful. The
estimated relaxation time for charm production and annihilation in
HG or even in a deconfined medium appeared to be much larger than
the total lifetime of the thermal system. Therefore, it has been
generally believed that the production mechanism of heavy-flavored
hadrons is completely different from that of light ones. For
instance,  the standard picture of charmonium production in
nucleus-nucleus collisions assumes that charmonia are created
exclusively at the initial stage of the reaction in primary
nucleon-nucleon collisions. During the subsequent evolution of the
system, the number of hidden charm mesons is reduced because of
absorption of pre-resonance charmonium states in the nuclei
(the normal nuclear suppression), interactions of charmonia with
secondary hadrons (comovers), dissociation of $c\bar{c}$ bound
states in the deconfined medium. The last mechanism was expected
to be especially strong and charmonia were proposed to be used 
as a probe of the state of matter created at the early
stage of the collision. It was found that the $J/\psi$ suppression
with respect to Drell-Yan muon pairs measured in proton-nucleus
and nucleus-nucleus collisions with light projectiles can be
explained by the normal nuclear suppression alone\cite{NA38}. In
contrast, the NA50 experiment with a heavy projectile and target
(lead-lead) revealed essentially stronger $J/\psi$ suppression for
central collisions\cite{anomalous}. This {\it anomalous} $J/\psi$
suppression was attributed\cite{evidence} to formation of quark-gluon 
plasma (QGP). In the same time, purely hadronic scenarios still
cannot be excluded\cite{comover}. Despite of quite successful 
agreement with the
$J/\psi$ data, the standard scenario seems to be in trouble
explaining the $\psi'$ yield. The recent lattice 
simulations\cite{Karsch} suggest that the temperature of $\psi'$ 
dissociation $T_d(\psi')$ lies far below the deconfinement 
point\cite{Satz01} 
$T_c$: $T_d(\psi') \approx 0.1$--$0.2 T_c$. Therefore, not
only the quark-gluon plasma, but also a hadronic co-mover medium
should completely eliminate $\psi'$ charmonia in central Pb+Pb
collisions at SPS. However, the experiment revealed a sizable
$\psi'$ yield (see, for instance\cite{Bordalo}). It was 
observed\cite{Shuryak} that $\psi'$ to $J/\psi$ ratio decreases with
centrality only in peripheral lead-lead collisions, but saturates
at sufficiently large number of participants $N_p \ge 100$. The
value of the ratio in (semi)central collisions is approximately
constant and equal to the ratio of the densities of these charmonium
states in an equilibrium HG. Hence, on one hand, the production of
charm cannot be thermal, because of the large relaxation time, on
the other hand, the multiplicity ratio of different charmonium
states is thermal. This paradox is resolved in the statistical
coalescence model (SCM)\cite{BMS}. In this model, the charm
quarks $c$ and antiquarks $\bar{c}$ are created at the initial
stage of A+A reaction in the hard parton collisions. This is
similar to the standard approach. But like in the thermal model\cite{GG}, 
the formation of observed hadrons with open and hidden
charm takes place later at the hadronization stage near the point
of chemical freeze-out. 
Production and annihilation of charm
quark-antiquark pairs at all stages after the initial are
neglected. Therefore the number $c$ and $\bar{c}$ in the system
may be very far from chemical equilibrium, but their distribution
over different species of hadrons with open and hidden charm is
controlled by the lows of statistical mechanics.

In the `pure' SCM, which will be a subject of the present discussion, 
it is assumed that the hot
strongly-interacting medium destroys all initially formed 
(`primordial') charmonia or even prevents their formation.
A combined scenario, assuming that some of primordial charmonia 
do survive suppression in the medium, but additional hidden 
charm meson can be as well formed at hadronization, has been also 
considered\cite{GR}.

Within the grand
canonical approach, the deviation of the amount of (anti)charm in
the system from its equilibrium value can be taken into account
simply by multiplying the multiplicities
of all open single (anti)charm species
in equilibrium hadron gas by some factor $\gamma_c$:
\begin{equation}\label{X}
\langle X \rangle = \gamma_c N_X(T,\mu_b,V) = \gamma_c 
\exp\left(c_X \frac{\mu_c}{T} \right) \tilde{N}_X(T,\mu_b,V).
\end{equation}
Here $c_X = \pm 1$ is the charm of the species $X$.
The charm chemical potential $\mu_c$ is needed to keep the total
balance of open charm and anticharm in the thermal system. 
At zero $\mu_c$ but nonzero baryonic chemical potential $\mu_b$, 
the total number of open charm in the equilibrium HG
\begin{equation}\label{N1}
\tilde{N}_1 = \sum_X \tilde{N}_X(T,\mu_b,V)
\end{equation}
is {\it not} equal to that of anticharm
\begin{equation}\label{N1b}
\tilde{N}_{\bar{1}} = \sum_{\bar{X}}
\tilde{N}_{\bar{X}}(T,\mu_b,V).
\end{equation}
This is because, for instance, a positive  value of $\mu_b$ enhances 
the number of
open charm baryons, that contribute to (\ref{N1}), and suppresses
the number of antibaryons in (\ref{N1b}). (The sums in (\ref{N1})
and (\ref{N1b}) run over all species of single open charm and
anticharm hadrons, respectively.) The value of $\mu_c$ should be
chosen to restore the balance:
\begin{equation}\label{balance}
\exp(\mu_c/T) \tilde{N}_1 = \exp(-\mu_c/T) \tilde{N}_{\bar{1}}.
\end{equation}

A hidden charm meson contains a charm quark as well as an antiquark.
Therefore, their thermal numbers are not influenced by the charm
chemical potential. To take into account the deviation from the
chemical equilibrium, one should multiply the thermal numbers of
charmonia by $\gamma_c^2$:
\begin{equation}\label{Y}
\langle Y \rangle = \gamma_c^2 \tilde{N}_Y(T,\mu_b,V).
\end{equation}

Provided that the average number of charm quarks and antiquarks
$\langle C \rangle = \langle \bar{C} \rangle$ in the system is
known, the factor $\gamma_c$ can be found from the
equation:
\begin{equation}\label{eqgam}
\langle C \rangle = \gamma_c \exp(\mu_c/T) \tilde{N}_1 + \gamma_c^2
\tilde{N}_H \ .
\end{equation}
Here $\tilde{N}_H$ is the total number of hidden charm particles
in the system at chemical equilibrium:
\begin{equation}\label{NH}
\tilde{N}_H = \sum_Y \tilde{N}_Y(T,\mu_b,V).
\end{equation}
(The sum runs over all charmonium states.) The thermal number of
hidden charm is much smaller than that of open charm. Therefore,
if the number of $c\bar{c}$ pairs in the system is not extremely
large (does not exceed the chemical equilibrium value by a
factor of several hundreds), the hidden charm term in
(\ref{eqgam}) can be neglected and $\gamma_c$ is simply given
by
\begin{equation}\label{gamc}
\gamma_c = \frac{\langle C \rangle}{\exp(\mu_c/T) \tilde{N}_1}.
\end{equation}

The grand canonical approach does not exactly correspond to the
physical reality. It keeps balance only between the {\it average}
quantities of charm and anticharm, while the partition function
includes also configurations with nonequal numbers of charm quarks
and antiquarks in the system. This does not influence the result,
if the average number of $c\bar{c}$ pairs in the system is large
$\langle C \rangle,\langle \bar{C} \rangle \gg 1$ . In the
opposite case\cite{We}, however, it should be taken into account that
strong interactions conserve charm (weak interactions can be
neglected), therefore the number $C$ of charm quarks in the system
is always {\it exactly} equal to the number $\bar{C}$ of
antiquarks.

Let us first consider the situation, when the numbers of charm
quarks and antiquarks are fixed (not necessary equal to each
other). In this case, the partition function for particles with
charm can be written as
\begin{equation}
Z_{C\bar{C}}(T,\mu_b,V) = \frac{\tilde{N}_1^{C}}{C !}
\frac{\tilde{N}_{\bar{1}}^{\bar{C}}}{\bar{C}!} + \tilde{N}_H
\frac{\tilde{N}_1^{C-1}}{(C-1) !}
\frac{\tilde{N}_{\bar{1}}^{\bar{C}-1}}{(\bar{C}-1)!} + \dots
\end{equation}
The dots stand for terms corresponding to more than one hidden
charm meson in the system and also for terms including double and
triple charm baryons. These terms, as it was already mentioned
above, can be safely neglected, unless $C$ and $\bar{C}$ are
extremely large.

As far as $\tilde{N}_H \ll \tilde{N}_1 \tilde{N}_{\bar{1}}$ (which
is true in all practically interesting situations), the numbers of
open charm and anticharm hadrons are approximately equal to $C$
and $\bar{C}$, respectively, i.e. almost all charm quarks and
antiquarks hadronize into open charm particles. Only a very small
fraction form charmonia. The total number of hidden charm can be
easily calculated:
\begin{equation}\label{HCC}
\langle H \rangle_{C\bar{C}} = \tilde{N}_H \frac{\partial \log
Z_{C\bar{C}}}{\partial \tilde{N}_H} \approx C\bar{C}
\frac{\tilde{N}_H}{\tilde{N}_1 \tilde{N}_{\bar{1}}}
\end{equation}

In reality, however, the number $K$ of $c\bar{c}$ pairs is not fixed. It
fluctuates from one event to another. The pairs are
produced in independent collisions of nucleons whose  
number is large and the 
probability to produce a $c\bar{c}$ pair in  a single 
collision is small. Therefore, the
fluctuations should follow the Poisson law:
\begin{equation}\label{pois}
P(C=\bar{C}=K) =
 \exp(- \langle c\bar{c} \rangle_{AB(b)})
 \frac{\left( \langle c\bar{c} \rangle_{AB(b)}\right)^{K}}{K!},
\end{equation}
where $\langle c\bar{c} \rangle_{AB(b)}$ stands for average number 
of $c\bar{c}$ pairs produced by nuclei A and B colliding  at impact 
parameter $b$.

The number of hidden charm averaged over all events at fixed
centrality is given by the convolution of Eq.(\ref{HCC}) with the
probability (\ref{pois}). The result reads:
\begin{equation}\label{HAB}
\langle H \rangle_{AB(b)} = \langle c\bar{c} \rangle_{AB(b)}
(\langle c\bar{c} \rangle_{AB(b)}+1)
\frac{\tilde{N}_H}{\tilde{N}_1 \tilde{N}_{\bar{1}}}\ .
\end{equation}

If one interested in the multiplicity of a particular charmonium
species, one should replace the $\tilde{N}_H$ in the numerator of
(\ref{HAB}) by the thermal multiplicity of this species. It must be
also taken into account that an additional contribution to
multiplicities of low-lying charmonium states comes from decays of excited
ones.

Note that the ratios of the numbers of different charmonium states in
SCM are exactly the same as in equilibrium HG:
\begin{equation}\label{chrat}
\frac{\langle \psi' \rangle_{AB(b)}}{\langle J/\psi
\rangle_{AB(b)}} = \frac{\tilde{N}_{\psi'}}{\tilde{N}_{J/\psi}}.
\end{equation}
It agrees with the behavior of $\psi'$ to $J/\psi$ ratio in central
Pb+Pb collisions at CERN SPS. This is not the case for peripheral events
($N_p < 100$). It can be explained in the following way: SCM
is valid if the momenta of charm (anti)quarks (not their number!)
are thermalized. This is possible only in a sufficiently large system.

\begin{figure}[th]
\centerline{\epsfxsize=3.9in\epsfbox{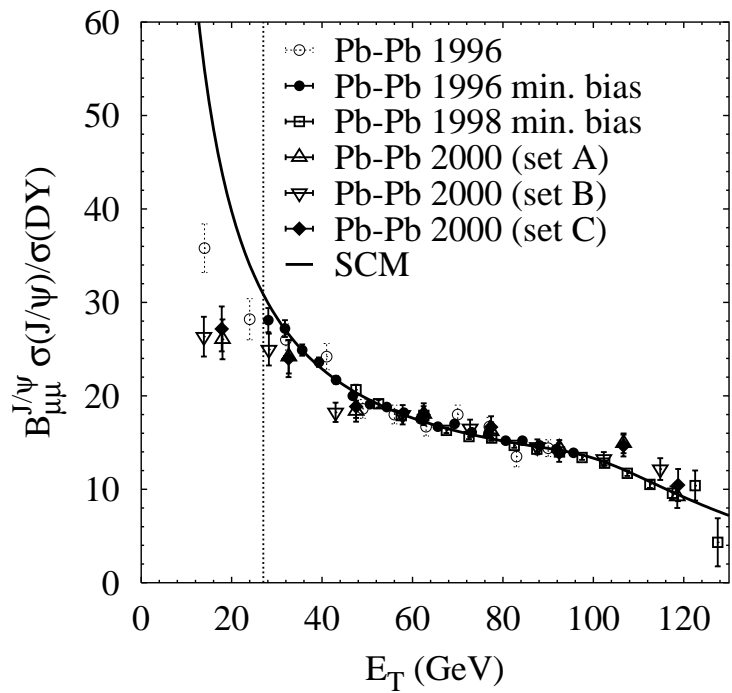}}   
\caption{The dependence of the $J/\psi$ to Drell-Yan
ratio on the transverse energy at SPS. The vertical line shows the
boundary of the
applicability domain of the statistical
coalescence model (SCM).}
\label{Jpsi_PbPb}
\end{figure}

The equation (\ref{HAB}) gives the total ($4\pi$) multiplicity of hidden
charm. In real experimental situation, however, measurements are
made in a limited rapidity window. In the most simple case, when
the fraction of charmonia that fall into the relevant rapidity
window does not depend on the centrality, one can merely use
Eq.(\ref{HAB}) multiplied by some factor $\xi < 1$. This situation
is likely to be relevant to charmonium production at CERN SPS,
where the multiplicity of light hadrons, that determine the
freeze-out volume of the system, are approximately proportional
to the number of nucleon participants $N_p$ at all rapidities.
Indeed, SCM fits well the measured\cite{anomalous,evidence,Ramelo}
$J/\psi$ to Drell-Yan ratio at CERN SPS
(see Fig. \ref{Jpsi_PbPb}, the details of the fit can be found in
Ref.\cite{SPS}).

At BNL RHIC, the situation is different: the total ($4\pi$)
multiplicity of light hadrons are still approximately proportional
to the number of participants, while at midrapidity, it 
grows faster\cite{PHOBOS}.
The centrality dependence of charmonium production
at different rapidities should in this case be also different. The formula 
(\ref{HAB}) should be generalized, in such a way 
that the charmonium multiplicity 
not only in the whole system, but in a part of it, a 
subsystem (like a limited rapidity interval), could be calculated.

Let, if $c$ (or $\bar{c}$) is present in the system,
the probability to find it in the subsystem
is $\xi < 1$. Then, if the number of $c\bar{c}$ pairs in the 
whole system is $K$, the probability to find $C$ charm quarks
in the subsystem is given by the binomial law:
\begin{equation}\label{binom}
w(C | K) = \frac{K!}{C! (K - C)!} \xi^C (1-\xi)^{K-C}.
\end{equation}
We assume that the distributions of quarks and antiquarks are 
uncorrelated,\footnote{At this point, our approach differs from that 
of Ref.\cite{andr}, where a strong correlation $C=\bar{C}$ between
the numbers of $c$-quarks and $\bar{c}$-antiquarks is assumed.} 
and the probability distribution of the number
of antiquarks $\bar{C}$ is given by the same binomial law. 

If the number of pairs $K$ is distributed in accordance with the 
Poisson law (\ref{pois}), the average multiplicity of charmonia in the 
subsystem is
\begin{eqnarray}  \label{Jpsi1}
\langle H \rangle_{AB(b)}^{(\xi)} &\approx &
\sum_{K=0}^{\infty} P(K)
\sum_{C=0}^{K} w(C | K)
\sum_{\bar{C}=0}^{K} w(\bar{C} | K) \ 
C \bar{C} \frac{\tilde{N}_H}{\tilde{N}_1 \tilde{N}_{\bar{1}}} \\
&=&
\xi^2 \ \langle c\bar{c} \rangle_{AB(b)} \ 
(\langle c\bar{c} \rangle_{AB(b)}+1) \ 
\frac{\tilde{N}_H}{\tilde{N}_1 \tilde{N}_{\bar{1}}}.
\nonumber
\end{eqnarray}
Here the meaning of $\langle c\bar{c} \rangle_{AB(b)}$ is the same as 
in Eq.(\ref{HAB}): the average {\it total} (4$\pi$) number of 
$c\bar{c}$ pairs in an event with given centrality. In contrast,
$\tilde{N}_H$, $\tilde{N}_1$ and  $\tilde{N}_{\bar{1}}$ are related 
to the subsystem. These are the thermal multiplicities 
of charmonia, open charm and anticharm calculated with the thermodynamic
parameters fitted within the HG model 
to the chemical composition and multiplicity of light 
hadrons {\it in the rapidity window under interest}.

\begin{figure}[ht]
\centerline{\epsfxsize=3.9in\epsfbox{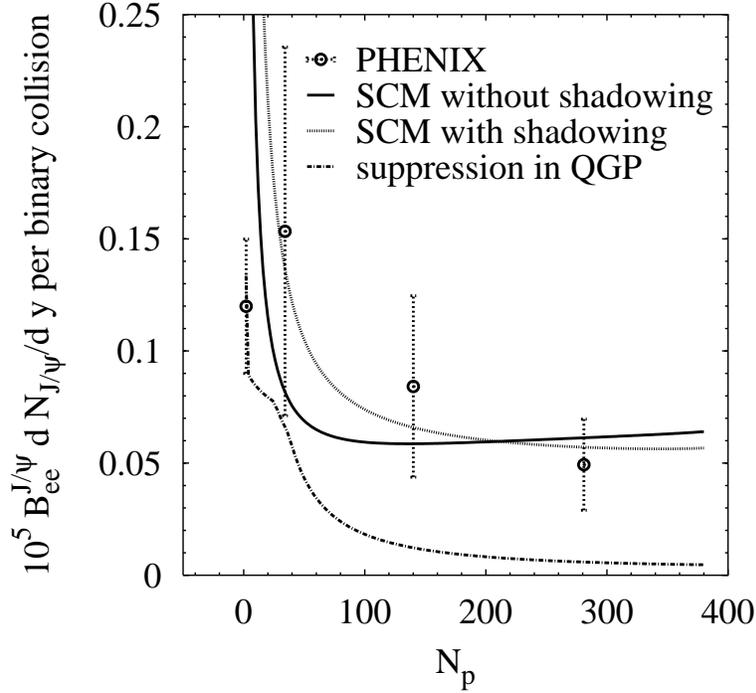}}   
\caption{The $J/\psi$ multiplicity 
per binary collision in the central unit rapidity interval
at $\sqrt{s}=200$ GeV muliplyed by the $J/\psi \rightarrow e^+ e^-$
branching ratio versus the number of nucleon participants. \label{Jpsi_RHIC}}
\end{figure}

In Fig.\ref{Jpsi_RHIC}, the SCM result is compared to the 
preliminary PHENIX data\cite{data} for $J/\psi$ production in the central 
rapidity interval at the top RHIC energy ($\sqrt{s}=200$ GeV).
The values of freeze-out parameters were taken from Ref.\cite{BMSR}:
$T=177$~MeV and $\mu_{B}=29$~MeV. The dependence of volume on the 
centrality was chosen to reproduce the measured multiplicity 
of charged hadrons.\cite{PHOBOS}
The value of $\langle c\bar{c} \rangle_{AB(b)}$ was calculated 
in the Glauber approach. The charm production cross section 
in nucleon-nucleon collisions was fixed at\cite{Averbeck}
$\sigma^{NN}_{c\bar{c}} = 650\ \mu$b, while $\xi$ was considered 
as a free parameter. 
The minimum value of $\chi^2/dof=0.75$ is 
reached at  $\xi \approx 0.18$. 

As is seen, the SCM dependence of the $J/\psi$ 
multiplicity {\it at midrapidity}
per binary collision on the centrality is almost flat
for $N_p \gtrsim 100$ in contrast to the  {\it total} 
$J/\psi$ yield , which is expected to 
grow in the same centrality region.\cite{We_RHIC}
This is because the multiplicity of light hadrons\cite{PHOBOS} 
and, consequently, the hadronization volume {\it at midrapidity} 
grows faster with the centrality than the {\it total} volume.
If the influence of the in-nuclear modification of the parton distribution 
functions (shadowing) on charm production at RHIC is essential,
a decrease should be observed instead of saturation
(see Fig.\ref{Jpsi_RHIC}).

The centrality dependence of $J/\psi$ multiplicity 
in the QGP suppression scenario\cite{Blaizot} is also shown in 
Fig.\ref{Jpsi_RHIC}. The parameters of the model were fixed from SPS 
data\cite{anomalous,evidence,Ramelo} and then extrapollated to RHIC.
As is seen, the observed yield can hardly be explained by survived 
primordial charmonia only. It seems that $c\bar{c}$ coalescence
at late stages of the reaction is here a dominant if not the 
only charmonium production mechanism.   

In conclusion, the statistical coalescence model is consistent 
with the data on $J/\psi$ production in heavy nucleus collisions
at SPS and RHIC energies, while the standard scenario of 
$J/\psi$ dissociation in QGP predicts too strong suppression at RHIC
that is not favored by the experimental data.


\section*{Acknowledgment}
The author is thankful to the Alexander von Humboldt Foundation
(Germany) for financial support.

\end{document}